\documentclass[twocolumn,lineno,dvipsnames, usenames]{aastex63}

\RequirePackage{lineno}

\usepackage{graphicx}	
\usepackage{amsmath}	
\usepackage{amssymb}	
\usepackage{enumitem}
\usepackage{hyperref}
\hypersetup{
    colorlinks=true,
    linkcolor=blue,
    filecolor=magenta,      
    urlcolor=cyan,
}

\newcommand{\Msun}{\,{\rm M_\odot}}
\newcommand{\Mblack}{m_\bullet}

\newcommand\mybar{\kern1pt\rule[-\dp\strutbox]{.8pt}{\baselineskip}\kern1pt}
\newcommand{\Ratio}{{\cal R}}
\newcommand{\Prob}{{\cal P}}
\newcommand{\BL}{\texttt{Baseline }} 
\newcommand{\ND}{\texttt{Stunted }}

\setlist[itemize]{noitemsep, topsep=0pt, leftmargin=*}

\shorttitle{Seeds problem with multi-band gravitational-wave observatories}
\shortauthors{Chen, Ricarte \& Pacucci}

\begin{document}

\title{Prospects to Explore High-redshift Black Hole Formation \\ with Multi-band Gravitational Waves Observatories}

\correspondingauthor{Hsin-Yu Chen}
\email{himjiu@mit.edu}

\author[0000-0001-5403-3762]{Hsin-Yu Chen}
\affil{Department of Physics and Kavli Institute for Astrophysics and Space Research, Massachusetts Institute of Technology, 77 Massachusetts Ave, Cambridge, MA
02139, USA}
\affil{LIGO Laboratory, Massachusetts Institute of Technology, 185 Albany St, Cambridge, MA 02139, USA}

\author[0000-0001-9879-7780]{Angelo Ricarte}
\affil{Black Hole Initiative, Harvard University,
Cambridge, MA 02138, USA}
\affil{Center for Astrophysics $\vert$ Harvard \& Smithsonian,
Cambridge, MA 02138, USA}

\author[0000-0001-9879-7780]{Fabio Pacucci}
\affil{Black Hole Initiative, Harvard University,
Cambridge, MA 02138, USA}
\affil{Center for Astrophysics $\vert$ Harvard \& Smithsonian,
Cambridge, MA 02138, USA}

\begin{abstract}
The assembly of massive black holes in the early universe remains a poorly constrained open question in astrophysics. The merger and accretion of light seeds (remnants of Population III stars with mass below $\sim 1000\Msun$) or heavy seeds (in the mass range $10^4-10^6 \Msun$) could both explain the formation of massive black holes, but the abundance of seeds and their merging mechanism are highly uncertain. In the next decades, the gravitational-wave observatories coming online are expected to observe 
very high-redshift mergers, shedding light on the seeding of the first black holes. In this Letter we explore the potential and limitations for LISA, Cosmic Explorer and Einstein Telescope to constrain the mixture ratio of light and heavy seeds as well as the  probability that central black holes in merging galaxies merge as well.
Since the third generation ground-based gravitational-wave detectors will only observe light seed mergers, we demonstrate two scenarios in which the inference of the seed mixture ratio and merging probability can be limited. The synergy of multi-band gravitational-wave observations and electromagnetic observations will likely be necessary in order to fully characterize the process of high-redshift black hole formation. 

\end{abstract}


\section{Introduction}

Most massive galaxies host a super-massive black hole ($\Mblack \gtrsim 10^6 \Msun$) at their centers, and significant correlations between its mass and some physical properties of the host galaxy exist \citep{Ferrarese_Merritt_2000, Gebhardt_2000, McConnell_Ma_2013}, indicating a strong case for their co-evolution \citep{Kormendy_Ho_2013}. While these correlations have been ascertained by decades of studies, various aspects of the formation mechanism of central black holes (BHs) are still unclear. BHs can grow by accreting mass and/or by merging with each other, with each mechanism more or less likely to occur as a function of its mass and of the cosmic time \citep{Pacucci_2020}. Several pathways to form massive BHs in early (i.e., $z \gtrsim 15$) galaxies have been put forward (see, e.g., reviews by \citealt{Woods_2019, Inayoshi_Haiman_2020}). The most straightforward mechanism to form BHs would require seeds formed as remnants of Population III stars \citep{Woosley_2002, Hirano_2014}, typically with a mass $\Mblack \lesssim 10^3 \Msun$: these are commonly referred to as \emph{light seeds} (see, e.g., \citealt{Natarajan_2011}). Alternatively, \emph{heavy seeds} with typical mass $\Mblack =10^4-10^6 \Msun$ could have formed by exotic pathways at $z > 10$ \citep{Bromm_Loeb_2003, Lodato_Natarajan_2006}. While additional, alternative pathways to form massive BHs at high redshift have been described in the literature (see, e.g., \citealt{Devecchi_2009, Davies_2011}), including medium-weight seeds \citep{Sassano_2021}, the classic distinction between light and heavy seeds will be considered here for the sake of clarity. 
To date, it is still unknown whether light seeds, heavy seeds or, more likely, a combination of them contributed to the formation of the population of the first BHs. 

Several studies explored the prospects of differentiating heavy seed from light seed models with gravitational wave (GW) detections (e.g., \citealt{Klein_2016, Ricarte_2018_seeds, Pacucci_2020, Valiante_2021, Toubiana_2021}), but the actual scenario can be far from binomial. Various mixtures of light and heavy seeds could have originated at high-redshift, and it is still unclear the extent to which future observatories will be able to pinpoint the exact mixture ratio. 
From the electromagnetic side, efforts are underway to find observational smoking guns to pinpoint the mixture of light/heavy seeds in the high-redshift universe. Some studies (e.g., \citealt{Pacucci_2017, Natarajan_2017, Valiante_2018, Whalen_2020}) have explored the observational properties of heavy seeds and predicted that the \textit{James Webb Space Telescope} (JWST) will be able to observe them up to $z \sim 15$. On the contrary, light seeds seem to be too faint to be observed even with the JWST \citep{Natarajan_2017, Wang_2017, Pacucci_decadal}. Some studies (e.g., \citealt{Ricarte_2018_seeds}) suggest that the several probes (e.g., the luminosity function of high-$z$ quasars, as well as the mass function obtained via GW observations) could be used to constrain the mixture of light/heavy seeds.

A significant challenge in the determination of the mixture ratio of light/heavy seeds via GWs is related to our lack knowledge of the efficiency of the merging process, or, in other words, the probability that once a merger between two host galaxies occurs, also their central BHs will merge. The calculation of the BH merging probability is very complicated \citep{Klein_2016}. Often, semi-analytical models assume a fixed probability regardless of the physical properties of the merging halos for simplicity (e.g., \citealt{Ricarte_2018_samintro}). Some recent studies (e.g., \citealt{Tremmel_2018}) are presenting a more complex picture, suggesting that the merging probability does depend on the stellar mass of the two merging halos, at least in the case of super-massive BHs.  It is also possible for orbital decay to stall on parsec scales, when dynamical friction becomes inefficient \citep{Milosavljevic&Merritt2001,Milosavljevic&Merritt2003}.  The possibility of measuring the merging probability with future GW detectors is certainly tantalizing, because such measures would also provide constraints on the  merger mechanisms in action.

The GW detectors coming online in the next decades have great potential to shed light on the seeds of massive BHs, by directly observing their mergers at high redshifts. The Laser Interferometer Space Antenna (LISA), scheduled to launch in early 2030s, will be able to observe binary black hole mergers (BBHs) from few hundreds to $\sim 10^6 \Msun$ at $z>10$ \citep{LISA_2017}. On the other side of the GW spectrum, next-generation (or third- generation, 3G) ground-based GW detectors with improved designs are being actively planned and investigated. Among them, detectors with emphasis on low-frequency sensitivity, such as the Einstein Telescope (ET, \citealt{ET_2010}) and the Cosmic Explorer (CE, \citealt{CE_2019}), can detect BBHs up to a thousand solar masses in the early universe~\citep{2019CQGra..36v5002H}. The mass spectra covered by LISA and 3G detectors are complimentary to each other and will provide our main discovery window into the mergers of BHs in the next decades.

A full picture of BH seeding models can only be achieved by looking at very high redshifts, above $z\sim 10$. As BHs do not keep memory of their growth and merging history, the observation of lower-redshift merging events could mislead the investigation because, for example, BHs might have grown significantly from their seeding stage. Note that the cosmic time between $z \sim 10$ and $z\sim 7.6$ (the redshift of the farthest, full-grown quasar discovered so far, see \citealt{Wang_2021}) is $\sim 4$ times the quasar growth timescale, i.e., the Salpeter time. Fortunately, both LISA and CE/ET will be able to detect mergers at very high redshift, making them the perfect instruments to obtain a full picture of BH seeding at high redshift. 

Seen with the eyes of a GW detector, the number of detections at a specific mass scale and redshift will depend on both the abundance of seeds and their merging probability. LISA is mostly sensitive to the mergers of heavy seeds, while the focus of 3G is on the merger of light seeds and intermediate seeds. In this study, we explore the potential and limitations of LISA and 3G observations to probe the mixture ratio between light and heavy seeds, as well as their merging probability, in the early universe.

\section{Methods}\label{sec:methods}
In this Section we describe our models for BH seeding and growth, and how we infer the predicted number of BBHs observed by LISA and 3G detectors.

\subsection{Modeling BH Assembly}\label{subsec:seed_model}

We use as our starting point a model for BH assembly from $z=20$ to $z=0$ developed in \citet{Ricarte_2018_samintro,Ricarte_2018_seeds}\footnote{These models correspond to the ``Power Law'' light and heavy seed models explored in this previous work, but with ``$p_\mathrm{merge}=1$'' until different merger probabilities are probed in post-processing.}.  BH assembly is tied to that of galaxies, which is, in turn, tied to the assembly of dark matter halos.  Thus, dark matter merger trees form the backbone of this model, and these are computed using a binary Press-Schechter algorithm calibrated against the Millenium simulations \citep{Parkinson_2008}.  In this model, either ``light'' or ``heavy'' BHs are seeded in the redshift range of $z \in [15,20]$.  Then, as time progresses, the fundamental assumption of BH growth in this model is that (i) periods of Eddington-limited growth are triggered by major galaxy mergers and (ii) this growth period ceases when the BH reaches the $M_\bullet-\sigma$ relation, where $\sigma$ is the host galaxy's velocity dispersion.  This can be interpreted as either the galaxy only providing enough fuel to place a BH on this relationship, or BH feedback stopping growth once this limit is reached, as suggested by analytic models \citep[e.g.,][]{Haehnelt+1998,King2003}.  In order to focus our modeling and computational power on BHs, empirical relationships between dark matter halos and galaxy properties are used to estimate $\sigma$ of a given galaxy from stellar masses and effective radii \citep[][]{Moster+2013,Mosleh+2013,Larkin&McLaughlin2016}.  

When two dark matter halos merge, their BHs merge as well after a dynamical friction timescale elapses, estimated from the \citet{Boylan-Kolchin+2008} formula.  This prescription naturally produces some ``wandering'' BHs, which fail to merge within the age of the universe simply because their dynamical friction timescales are too long.  It is known that dynamical friction becomes inefficient at scales of roughly a parsec, potentially leading to what is referred to as the ``final parsec problem'' \citep[e.g.,][]{Milosavljevic&Merritt2001,Milosavljevic&Merritt2003}.  There are now many proposed solutions to this problem, including non-axisymmetry of the gravitational potential \citep{Khan+2013,Rantala+2017,Gualandris+2017} and the assistance of gas \citep{Ostriker1999,Escala+2004,Armitage&Natarajan2005,Mayer+2007,Fiacconi+2013,Goicovic+2017}.  In this work, we do not model the evolution of BBHs and assume that the ``final parsec problem'' is resolved on a timescale short compared to the dynamical friction timescale.

The assembly of BHs at early epochs is extremely uncertain and poorly constrained.  On the one hand, quasars at redshifts $z>6$ have been observed with masses already in excess of a billion solar masses, implying that at least some of the BH population must have grown very efficiently in the early universe.  Some objects require either super-Eddington accretion for most of the age of the universe to that point, or at least near-Eddington growth with a heavy seed \citep[e.g.,][]{Fan_2003,Volonteri_2005_quasars,Banados_2018}.  On the other hand, several galaxy-scale simulations point towards a picture in which supernova feedback inhibits the growth of early black holes, blowing away gas in shallow potentials before it has the opportunity to accrete onto BHs \citep{Angles-Alcazar+2017,Habouzit+2017,Bower+2017}. Hence, for each of our seeding scenarios, we explore two different BH growth models, one where BH growth is inhibited by supernova feedback (\ND model), and another where it is not (\BL model).

In our \BL model, major mergers with halo mass ratios of 1:10 or larger trigger BH growth at the Eddington rate until the BH reaches the $M_\bullet-\sigma$ relation.  In the \ND model, we impose the additional condition that the host halo must be above a critical halo mass $M_\mathrm{crit} = 10^{12}\Delta_z^{-3/8} \; M_\odot$, where $\Delta_z = [\Omega_m(1+z)^3 + \Omega_\Lambda]^{1/3}$.  Above this mass, outflows from supernova feedback cease to be buoyant in an analytic model \citep{Bower+2017}.  \citet{Barausse_2020} found that BH growth-inhibiting supernova feedback could decrease LISA events.  However, as we shall show, the same physical mechanism can increase the number of mergers detectable by CE/ET by causing light seeds to stay in the detectable mass range for a longer time.  
 
\subsection{Mixture ratio}\label{sec:MT}
Due to significantly more stringent physical requirements, the formation of a heavy seed is always less likely than the formation of a light seed (see, e.g., \citealt{Latif_2013b, Habouzit_2016}). Similarly to \citet{Toubiana_2021} we probe the effect of a variable mixture ratio of light vs heavy seeds in a post-processing step.  We parametrize with $\Ratio$ the probability of forming a light seed, whenever the physical conditions of the host halo allow the formation of both of a light and of a heavy seed. It follows that the probability of forming a heavy seed is $1- \Ratio$.  Note that this occurs in post-processing, and implicitly, a merger between light and heavy seed cannot occur.

\subsection{Merging probability}\label{sec:probability}

As discussed in \autoref{subsec:seed_model}, the probability that two central BHs merge as a consequence of a galaxy merger is highly uncertain, with uncertainties ranging from kiloparsec to parsec scales.  
Properties such as the mass of the BH pair, as well as the stellar mass of the merging galaxies and their gas content can strongly influence the merging time and, indeed, the possibility that the two BHs merge at all.  This merging probability is, as a matter of fact, the most important unknown folded in the calculation of high-$z$ merger rates.

On kiloparsec scales, dynamical friction is the dominant process by which BHs lose angular momentum \citep{Chandra_1943,Begelman_1980}.  However, in a realistic cosmological environment, \cite{Tremmel_2018} find that the close-pair formation timescale can be longer than expected, possibly even in excess of the Hubble time at the redshift when the merger of their hosts occurs.   Remarkably, the two central BHs may not even end up merging at all, resulting in a ``wandering'' population \citep[e.g.,][]{Reines_2020,Ricarte_2021,Weller_2022}. This problem can potentially be exacerbated in clumpy high-redshift galaxies.  On parsec scales, dynamical friction becomes inefficient, and additional processes are required to further shrink the BBH into the gravitational wave regime \citep[e.g.,][]{Colpi2014}.  Although many proposed solutions exist to this problem (see \autoref{subsec:seed_model}), it remains unclear which process dominates, how significant delays might be, and how conditions may change in messy high-redshift environments.

For simplicity, here we use a merging probability factor $\Prob$ that is BH mass-independent, in accordance with previous studies.  Note that we already do include a sinking time dependence on the host halo mass following a halo merger in the merger tree, which could potentially exceed the Hubble time.  For the mass range most relevant for our studies, the merging probability is most likely driven not by the BH's individual mass, but rather the combined mass of the baryons in its vicinity, which is unknown in our model.  In cosmological simulations, \citet{Weller_2022} find that the sinking times of low-mass BH proxies in the IllustrisTNG simulations were not dependent on their individual masses, while \citet{Tremmel_2018} report a dependence on the central density of stars around the BH.  While a merging probability that is mass-independent will naturally under count and over count mergers at the extremes of the mass spectrum, any choice of a mass dependence over a very wide mass range ($\sim 10^1-10^7 \Msun$) would be arbitrary.

This assumption is made in light of the final goal of this study, which is to explore the possible degeneracy between light/heavy seed mixture ratio $\Ratio$ and the overall merger probability $\Prob$.

\subsection{LISA and 3G observations}
\label{subsec:obs}
The calculation of the signal-to-noise ratios for LISA detections is performed with the scripts provided by ~\citet{2019CQGra..36j5011R}.
For each BBH in our catalogues, we randomly assign: (i) sky location, (ii) inclination, and (iii) polarization angles. For a detection we require the signal-to-noise ratios of the BBHs to be larger than 8 (see also \citealt{LISA_2017}).

For 3G ground-based detectors, we follow ~\cite{Chen:2017wpg} to estimate the fraction of detectable BBHs after averaging over sky locations, inclination and polarization angles. We use the official \href{https://dcc.ligo.org/LIGO-T1500293/public}{sensitivity curves} for CE (file \texttt{ce2.txt}) and ET (file \texttt{et\_d.txt}). Our calculations consider a single detector measurement and the detection threshold is set at signal-to-noise ratio of 8. 
We note that increasing the number of detectors at different locations will increase the total number of detections, but the detectable mass and redshift are not expected to change significantly.
 
For both LISA and 3G detectors, we assume four years of observations. Since we simulated BBHs in their source frame, a $1/(1+z)$ factor was applied in the estimate of the number of detections in order to take into account the time dilation due to cosmic expansion. Planck 2015 cosmology~\citep{Planck_2015} cosmology was used throughout the simulations and calculations.

\section{Results}\label{sec:results}
\begin{figure*}
    \centering
    \includegraphics[width=0.8\linewidth]{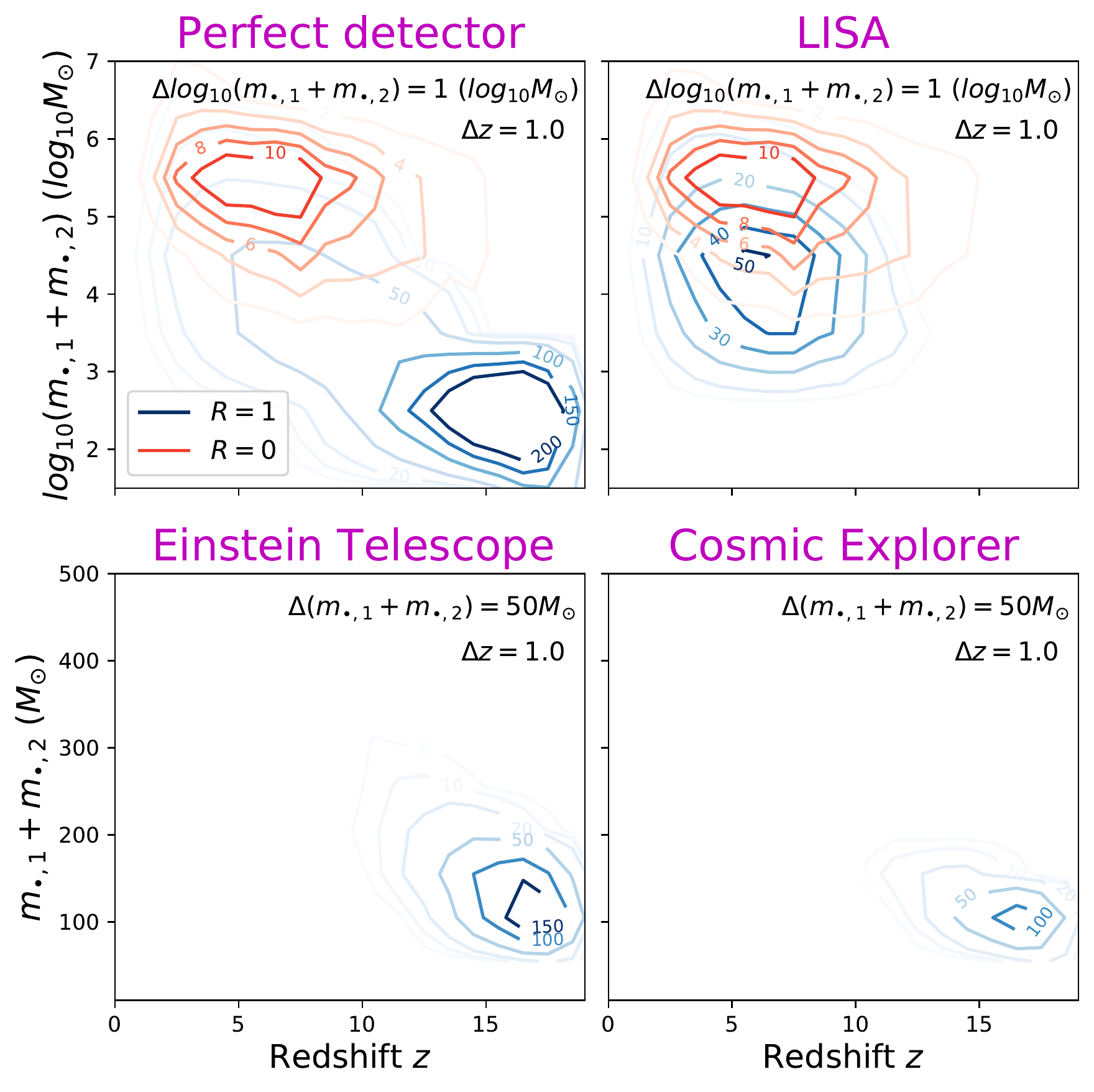}
    \caption{Number of detections binned by the redshift and total mass of BBHs for a perfect detector, LISA, Einstein Telescope, and Cosmic Explorer over four years of observations. The bin sizes are indicated in each subfigure. The labels of the contours indicate the number of detections in four years for any given redshift and mass bin. Two scenarios are presented in this figure: light-seed-only (blue, $\Ratio=1$) and heavy-seed-only (red, $\Ratio=0$). For both scenarios, the merging probability is $\Prob=1$. 
    }
    \label{fig:all_originalp}
\end{figure*}

We first present the total number of detections over four years of observations in the \BL model by LISA, ET, and CE assuming the BH formation were entirely dominated by light seed (blue) or heavy seed (red) in Figure~\ref{fig:all_originalp}. The number of detections are binned by the redshift and total mass of the mergers. In order to present the original number of mergers in our simulation, we also show the number of detections in a perfect detector, i.e., the detector that can observe all mergers without any detection limit. As expected, the heavy-seed-only scenario only leads to detections in the LISA band. On the other hand, ET and CE can detect more mergers beyond redshift of 10 than LISA in the light-seed-only scenario.

\begin{figure*}
    \centering
    \includegraphics[width=0.8\linewidth]{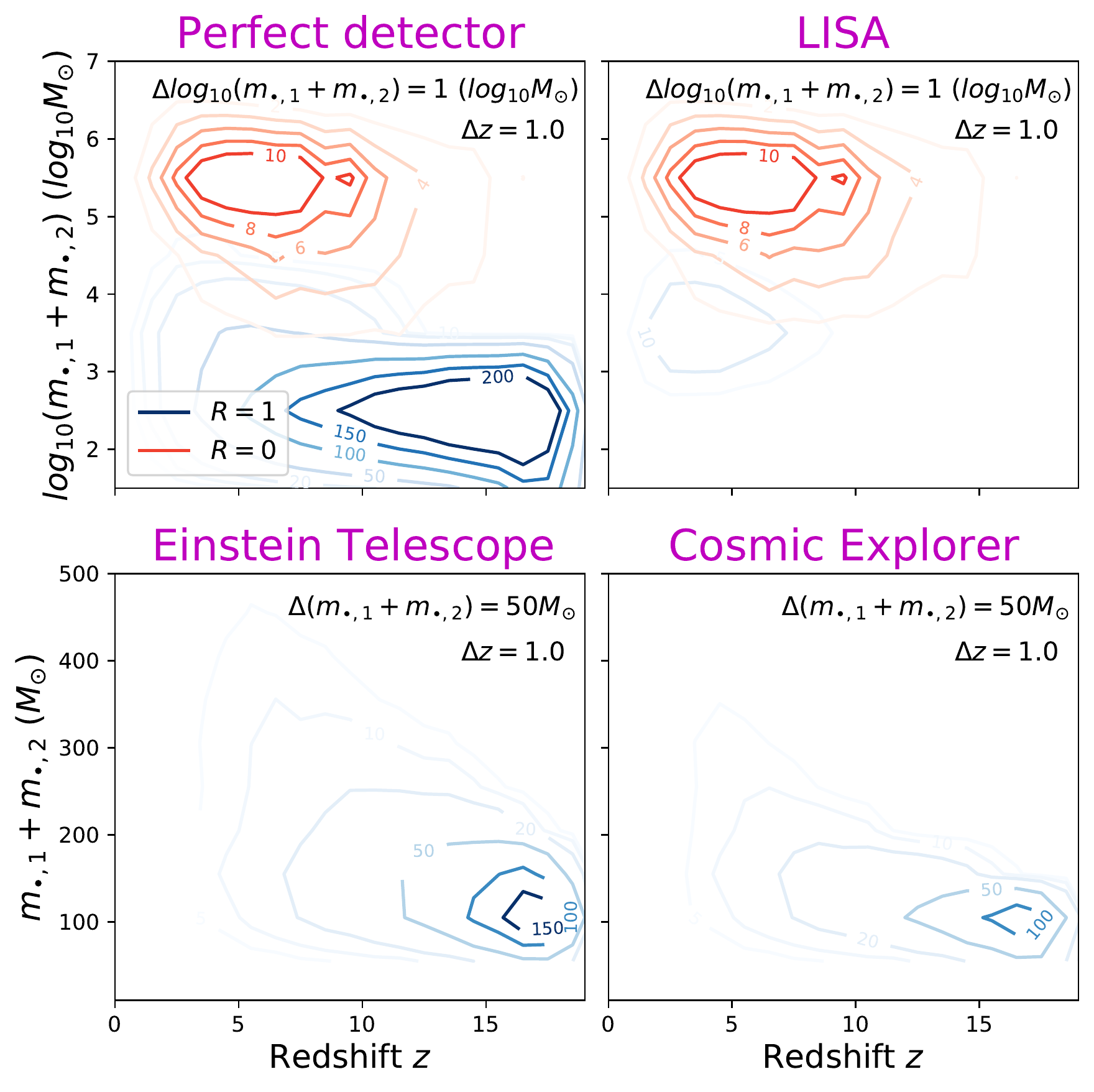}
    \caption{Same as Figure~\ref{fig:all_originalp} for the \ND model, inhibiting BH growth in low-mass halos due to supernova feedback allows for light seeds to remain in the detectable mass range for a longer period of time.}
    \label{fig:all_snbarrier}
\end{figure*}
Similarly, we show the total number of detections for the \ND model in Figure~\ref{fig:all_snbarrier}.  In this model, there are more light seed mergers detectable by CE/ET because their growth is stunted by supernova feedback and they do not grow out of the detectable mass range. Hence, we expect more low-mass BBHs at lower redshift in ET and CE in the light-seed-only scenario than in the \BL model. 

Next, we explore how the distributions of number of detections vary with the mixture ratio and merging probability. We notice distinct distributions among some seeding and merging probability scenarios, and therefore the GW detections are expected to produce informative constraints. For example, if it is known that the light seeds dominated (high $\Ratio$), the numbers of detections 
in both ground- and space-based detectors are directly proportional to the merging probability $\Prob$. The merging probability can then be inferred from the number of detections. However, in some other scenarios the limited access to heavier mergers by ET and CE introduce ambiguities. Although ET and CE are sensitive to mergers with total mass up to a thousand solar masses~\citep{2019CQGra..36v5002H}, the majority of detections will still lie below $\sim 200 \Msun$.  However, here we note that our model only accounts for a single nuclear BH in each galaxy and does not include stellar remnants that form at larger radius, outside of their centers.

With only masses and redshifts, we cannot observationally distinguish between light seed mergers at the centers of galaxies and ``normal'' mergers from BBHs that form throughout the galaxy at low masses, unless special features, such as spin properties, can be characterized theoretically and/or observationally in future developments. Since ordinary stellar evolution is expected to produce individual BH remnants only as massive as $\sim$90 $M_\odot$ \citep[e.g.,][]{Belczynski_2017,vanSon_2020}, we tentatively consider any BBHs with masses above $200 \Msun$ as distinguishable from this off-nuclear population formed through normal stellar evolution channels. Note that off-nuclear mergers above this mass limit can still occur as the result of second-generation mergers, although this channel is highly uncertain.
 
Below, we give two examples in which we expect limitations from ET and CE observations. 

\begin{figure}
    \centering
    \includegraphics[width=1.05\linewidth]{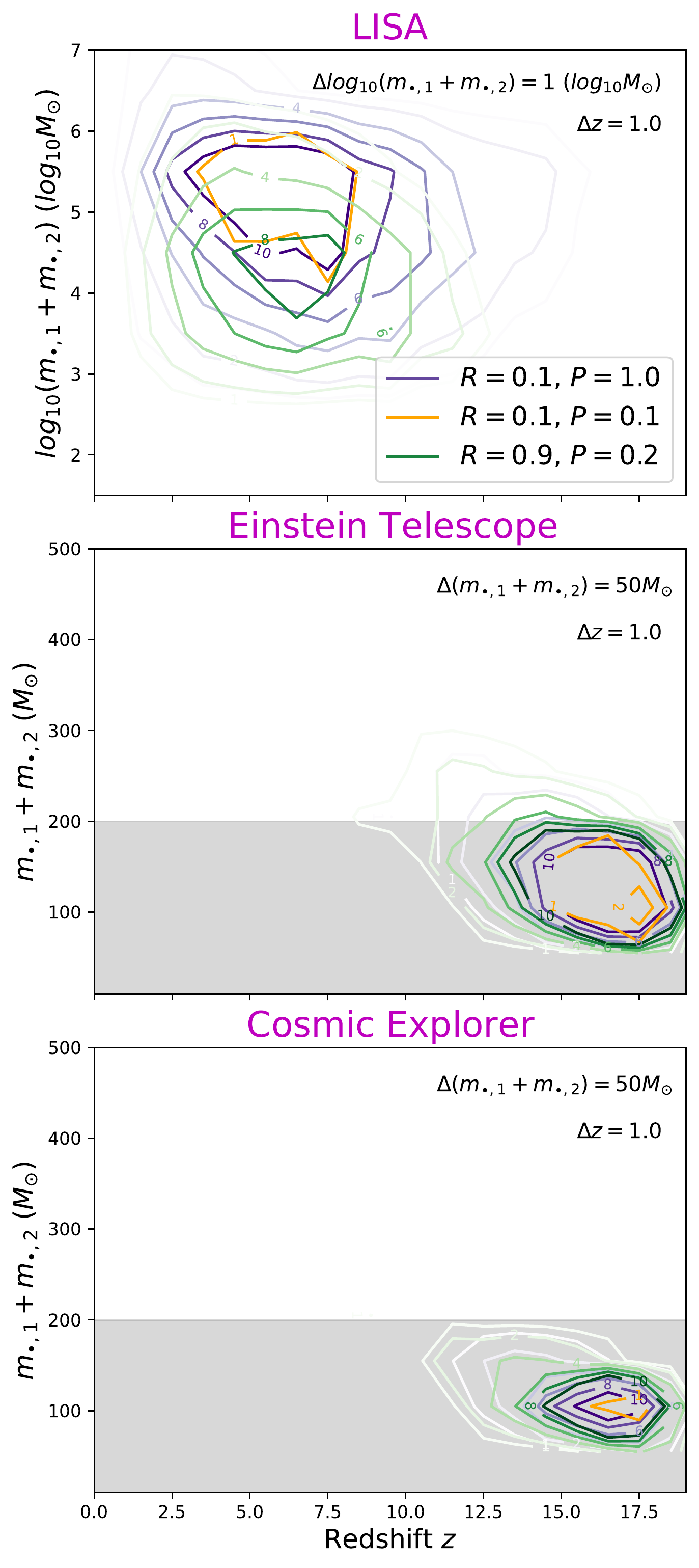}
    \caption{Same as Figure~\ref{fig:all_originalp} but with three different scenarios: heavy-seed-dominated and high merging probability ($\Ratio=0.1,\Prob=1.0$; purple contours), heavy-seed-dominated and low merging probability ($\Ratio=0.1,\Prob=0.1$; orange contours), and light-seed-dominated and low merging probability ($\Ratio=0.1,\Prob=1.0$; green contours). The grey bands approximately mark the mass range in which extra off-nuclear BBHs are expected and can lead to confusions in detections (see texts for more details).}
    \label{fig:RP_originalp}
\end{figure}

First, in the heavy-seed-dominated universe (low $\Ratio$), BBHs detections above $\sim 200 \, M_{\odot}$ by ET and CE are expected to be rare. Although the number of mergers is proportional to the merging probability, the difference in the number of detections may not be obvious between different values of $\Prob$. In Figure~\ref{fig:RP_originalp}, we compare the binned number of detections for $\Prob=0.1$ and $\Prob=1.0$ when $\Ratio=0.1$, assuming the \BL model. Although the distributions of detections are very different between these two scenarios, the discrepancies could be buried under the detections of off-nuclear populations in ET and CE.    

\begin{figure}
    \centering
    \includegraphics[width=1.05\linewidth]{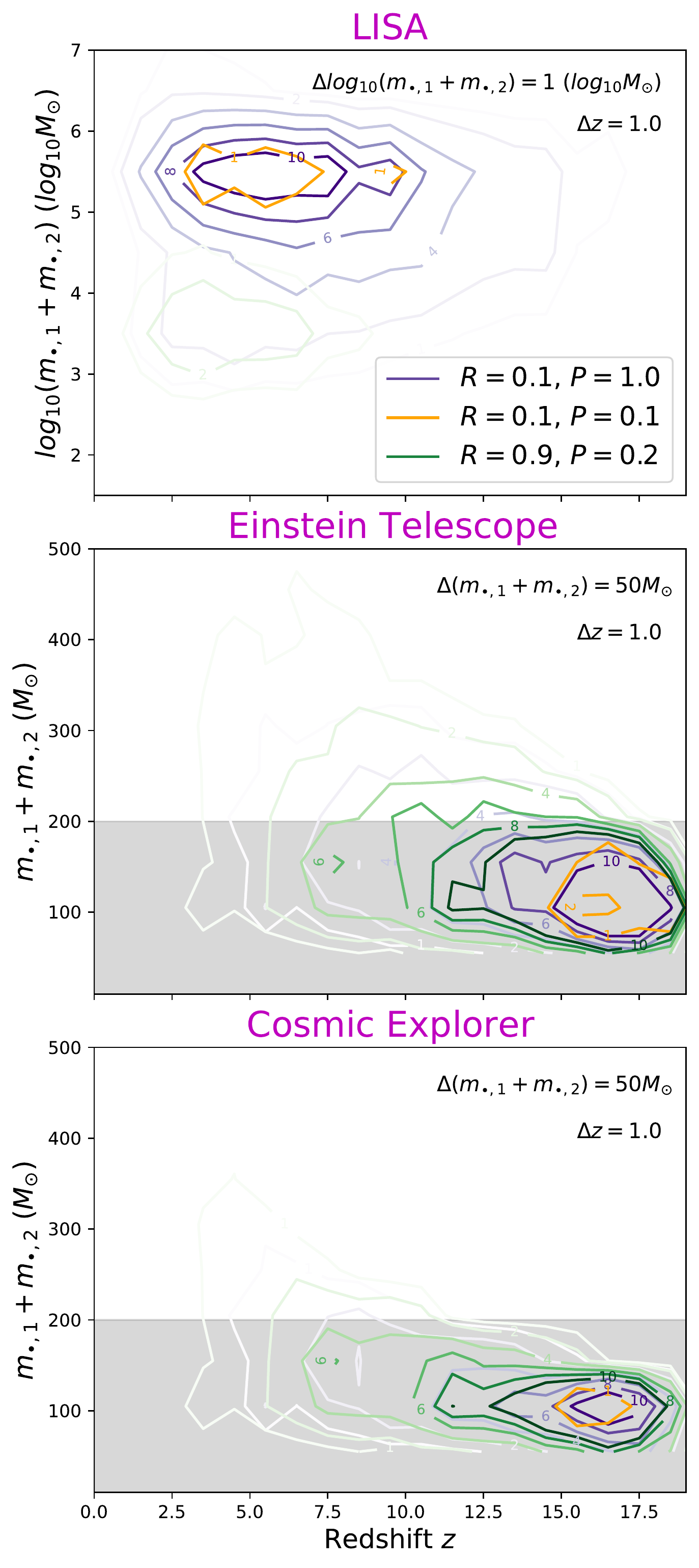}
    \caption{Same as Figure~\ref{fig:RP_originalp} for the \ND model.}
    \label{fig:RP_snbarrier}
\end{figure}
Second, a heavy-seed-dominated universe with high merging probability (low $\Ratio$ and high $\Prob$) can be indistinguishable from a light-seed-dominated universe with low merging probability (high $\Ratio$ and low $\Prob$). In Figure~\ref{fig:RP_originalp} we compare the scenario $(\Ratio=0.1, \Prob=1.0)$ to the scenario $(\Ratio=0.9,\Prob=0.2)$, assuming a \BL model. The distribution of detections for these two scenarios are similar in the ET and CE mass range, limiting the determination of the dominant population of seeds with the ground-based GW observations. We no longer observe this degeneracy with the \ND model, as shown in Figure~\ref{fig:RP_snbarrier}. The detections of BBHs at lower redshift ($z<10$) can serve as a smoking gun for a light-seed-dominated universe in this model, as they are limited to higher redshift for a heavy-seed-dominated universe. However, this feature at low redshift for a light-seed-dominated universe may still be indistinguishable if the off-nuclear populations can not be separated from the detections.

On the other hand, LISA observations of heavy BBHs could help differentiating the scenarios in the above two examples. In the first example, the distribution of detections in redshift and mass space as well as the absolute number of detections could constrain the merging probability if the universe was dominated by heavy seeds. In the second example, the difference in the expected mass distribution, especially the detections of the heaviest BBHs ($>10^6 \Msun$), helps separating the light and heavy seed scenarios.

\section{Discussion and conclusion}
In this Letter we explore the potential and limitations for LISA and 3G ground-based GW observations to unravel the mixture ratio of light and heavy seeds as well as the merging probability for two BHs after their host galaxies merge. We present two scenarios in which these properties can be nontrivial to constrain from 3G detections: 
\begin{enumerate}
    \item In a heavy-seed-dominated universe, the difference in redshift-mass distribution of detections due to different merging probabilities may be obscured by the populations of off-nuclear BBHs.
    \item A light-seed-dominated universe with low merging probability may lead to similar redshift-mass distribution of detections to a heavy-seed-dominated universe with high merging probability.
\end{enumerate}
We have shown that if supernova feedback inhibits early BH growth, light seeds can be more accessible to 3G detectors, since they do not grow out of the detectable mass range as easily.  We verify that even in the more optimistic model for 3G detections in which light-seed numbers are boosted (the \ND model), the above degeneracy cannot be easily resolved.

Nevertheless, 3G ground-based GW detectors will provide unparalleled access to light seed BBHs~\citep{2021ApJ...913L...5N,2021arXiv210807276N}. When combining with the knowledge of heavy seeds from future electromagnetic observations (see, e.g., \citealt{PFVD_2015, Natarajan_2017, Ricarte_2018_seeds, Valiante_2018, Pacucci_decadal, Whalen_2020}), we expect the degeneracy between mixture ratio and merging probability can be better resolved. In addition, the BH spin measurement is potentially another useful information to help constraining the seed properties that requires follow-up developments.  In particular, growth via persistently aligned accretion disks can spin up nuclear BHs to a near maximal value, which could potentially allow them to be distinguished from the off-nuclear population \citep[e.g.,][]{Thorne_1974,Volonteri_2005}.  Finally, a better estimate of the ratio between nuclear and off-nuclear BBHs will be crucial to disentangle the population of light seeds in the 3G detections.

We note that the seed models we investigate are not expected to capture a complete narrative of the seed evolution. Nonetheless, the challenges for 3G detections we demonstrate with our models are expected in other seed models as well. This is because the undetermined mixture ratio and merging probability always dominate the uncertainty of the number of mergers. Both the mixture ratio and the merging probability can vary the number of mergers within a given mass range. Therefore, observations targeting a limited mass spectrum will have difficulties to jointly constrain the mixture ratio and merging probability from the number of detections.

In this Letter we only discuss the degeneracy in the distribution of number of detections, assuming the mass and redshift of the BBHs can be precisely measured. In reality, the precision of parameter estimation for high-redshift BBHs can be another important challenge to overcome~\citep{2009CQGra..26i4027A,2010MNRAS.401.2706P,2017PhRvD..95f4052V} before the seed model can be finalized. Therefore, the synergy of multi-band GW observations and electromagnetic observations may not be contingent but necessary.

\acknowledgments{
We acknowledge valuable feedback from Carl-Johan Haster, Michele Maggiore, Antonio Riotto, Salvatore Vitale, and Rainer Weiss. HYC is supported by NASA through NASA Hubble Fellowship grants No.\ HST-HF2-51452.001-A awarded by the Space Telescope Science Institute, which is operated by the Association of Universities for Research in Astronomy, Inc., for NASA, under contract NAS5-26555. A.R. acknowledges support from the National Science Foundation under Grant No. OISE 1743747.  F.P. acknowledges support from a Clay Fellowship administered by the Smithsonian Astrophysical Observatory. This work was also supported by the Black Hole Initiative at Harvard University, which is funded by grants from the John Templeton Foundation and the Gordon and Betty Moore Foundation.
}

\bibliography{ref}

\end{document}